\documentclass[11pt,reqno]{JHEP3}

\usepackage[latin1]{inputenc}
\usepackage[tbtags]{amsmath}
\usepackage{amsfonts}
\usepackage{amssymb}

\def\be {\begin{equation}}
\def\ee {\end{equation}}
\def\bs#1\es{\begin{split}#1\end{split}}
\def\ba#1\ea{\begin{align}#1\end{align}}
\def\bg#1\eg{\begin{gathered}#1\end{gathered}}
\def\bea{\begin{eqnarray}}
\def\eea{\end{eqnarray}}
\newcommand{\bay}{\begin{array}}
\newcommand{\eay}{\end{array}}

\def\c{\chi}
\def\d{\delta}
\def\e{\epsilon}
\def\vare{\varepsilon}

\def\F{\Phi}
\def\g{\gamma}
\def\h{\eta}

\def\l{\lambda}

\def\o{\omega}
\def\p{\psi}
\def\P{\Psi}

\def\z{\zeta}

\def\pr{\prime} 
\def\ppr{^{\prime \prime}} 
\def\pa{\partial}

\def\fr{\frac}
\def\dep{\delta_\epsilon^\prime}
\def\bc{\big|_{\pa \cM} }

\def\bls{\bigg [}
\def\brs{\bigg ]}
\def\cM{\mathcal{M}} 
 
\def\nn{\nonumber}
\def\lra{\leftrightarrow}
\def\qrq{\quad\Rightarrow\quad}

\def\pl{\text{\tiny{(+)}}}
\def\mi{\text{\tiny{(--)}}}
\def\plmi{\text{\tiny{($\pm$)}}}
\def\inD{\widehat D^\pr}

\newcommand{\dst}{\displaystyle}

\newcommand{\epbar}{\ov \ep}
\newcommand{\psibar}{\ov \psi}

\newcommand{\chibar}{\ov \chi}

\newcommand{\eps}{\varepsilon}
\newcommand{\ep}{\epsilon}
\newcommand{\la}{\lambda}
\newcommand{\da}{\delta}
\newcommand{\om}{\omega}
\newcommand{\Ga}{\Gamma}
\newcommand{\ga}{\gamma}

\newcommand{\rr}{\prime}
\newcommand{\ov}{\overline}
\newcommand{\wh}{\widehat}

\newcommand{\half}{\frac{1}{2}}
\newcommand{\qter}{\frac{1}{4}}
\newcommand{\mc}{\mathcal}

\numberwithin{equation}{section}
\numberwithin{figure}{section}

\title{\bf The supermultiplet of boundary conditions in supergravity}

\author{ Dmitry V.~Belyaev \\
Institute for Fundamental Theory, Department of Physics,\\
 University of Florida, Gainesville,  FL 32611, USA\\
E-mail: \email{belyaev@phys.ufl.edu}
}
\author{ Tom G. Pugh\\
The Blackett Laboratory, Imperial College, \\
Prince Consort Road, London, SW7 2BZ, UK\\
E-mail: \email{thomas.pugh08@imperial.ac.uk}
}

\date{\today}

\abstract{
Boundary conditions in supergravity on a manifold with boundary relate the bulk gravitino to the boundary supercurrent, and the normal derivative of the bulk metric to the boundary energy-momentum tensor. In the 3D $N=1$ setting, we show that these boundary conditions can be stated in a manifestly supersymmetric form. We identify the Extrinsic Curvature Tensor Multiplet, and show that boundary conditions set it equal to (a conjugate of) the boundary supercurrent multiplet. Extension of our results to higher-dimensional models (including the Randall-Sundrum and Horava-Witten scenarios) is discussed.
}


\preprint{}

\begin{document}

\section{Introduction}

Supersymmetric (susy) theories for systems with boundaries have been of great interest for some time \cite{h1,h2,h3,h4,h5,h6,h7,h8,h9,h10,h11,h12,h13}. The most notable examples of this are the 11D Horava-Witten construction \cite{Horava:1996ma}, also known as Heterotic M-Theory, and the 5D Randall-Sundrum scenario \cite{Randall:1999ee,Altendorfer:2000rr,Lukas:1998yy,Lukas:1998tt}. In these theories one begins by considering a bulk supergravity action and then proceeds to couple boundary-localized matter to it. The construction of a supersymmetric action, in these theories, is complicated by the fact that the bulk Lagrangian, which we usually refer to as being invariant under supersymmetry, in fact varies into a total derivative. The bulk action then varies into a surface term.

To produce an invariant action, when the effects of boundaries are considered, one therefore typically resorts to using certain boundary conditions (b.c.). These relate the bulk fields, which are restricted to the boundary, to the boundary-localized matter fields. The boundary action is then constructed in such a way as to cancel the surface term, after the b.c. have been imposed. Clearly, a key feature of these \emph{`susy with b.c.'} constructions is the b.c. themselves, as without them the bulk and boundary non-invariances are unable to be related and so will not cancel. 

The choice of boundary conditions available is subject to two constraints. 
Firstly, the b.c. must vary into each other under supersymmetry, which we describe by saying that they \emph{`form an orbit'} \cite{Lindstrom:2002mc}.  In other words, the b.c. must be expressible as a susy multiplet. In rigidly supersymmetric models, the multiplets (superfields) of b.c. have been identified e.g. in \cite{Belyaev:2005rs,Belyaev:2008xk}. 
Secondly, the b.c. must also be consistent with the variational principle, which makes the construction of supersymmetric bulk + boundary actions quite non-trivial, especially in supergravity \cite{Bagger:2002rw,Moss:2003bk,Moss:2004ck,Bagger:2004rr,Falkowski:2005fm,Belyaev:2005rt,Pugh:2010ii}. For this reason, in several studies on the subject, a semi-consistent approach has been adopted where the b.c. used for supersymmetry do not match those derived from the action \cite{Belyaev:2007aj}.

The fully consistent and constructive approach was presented in \cite{Belyaev:2007bg}. There it was shown, in a 3D setting, that it is possible to identify `co-dimension one' (boundary) supermultiplets of the bulk supergravity and matter multiplets \emph{without imposing any boundary conditions}, and the procedure for constructing bulk + boundary actions that are \emph{`susy without b.c.'} was described. This formulation relies on the existence of auxiliary fields needed to form the multiplets. It was demonstrated, however, that in certain cases the elimination of auxiliary fields yields actions that remain \emph{`susy without b.c.'} ~\footnote{
It is thus still an open question whether the 11D Horava-Witten model \cite{Horava:1996ma,Moss:2003bk,Moss:2004ck} allows a \emph{`susy without b.c.'} formulation. Some obstacles in achieving this in the similar 5D setup have been discussed in \cite{Belyaev:2005rt,Belyaev:2006jg}.
}
In general, however, the elimination of auxiliary fields mixes boundary conditions with supersymmetry \cite{Belyaev:2005rs}, and one ends up with a \emph{`susy with b.c.'} formulation as described above, but which is guaranteed to be consistent.

In this paper, we will consider the 3D $N=1$ equivalent of the Horava-Witten setup, and will work with the \emph{`susy without b.c.'} formalism of \cite{Belyaev:2007bg}. The b.c. in this case are still present but are simply implied by the variational principle rather than being necessary for supersymmetry. As is well-known, the (`natural' \cite{Courant,Barth:1984jb}) b.c. in supergravity relate the normal derivative of the bulk metric (i.e. the extrinsic curvature tensor) to the boundary energy-momentum tensor \cite{Israel:1966rt}, and the bulk gravitino to the boundary supercurrent \cite{Moss:2004ck,Belyaev:2007aj}. We will cast these b.c. in a manifestly supersymmetric form, in which they relate the Extrinsic Curvature Tensor Multiplet (ECTM), which we explicitly construct in this paper for the first time, to the boundary Super Current Multiplet (SCM), first introduced in \cite{Ferrara:1974pz}.

In section \ref{3DHetM}, we will set up our supersymmetric 3D bulk + 2D boundary system while reviewing the formalism of \cite{Belyaev:2007bg}. We will derive the field equations and boundary conditions as they follow from the variational principle, and pose the question of fitting them into multiplets. In section \ref{secECTM}, we will construct a multiplet that contains the extrinsic curvature tensor $K_{m n}$. The verification that this ECTM transforms as a standard 2D $N=(1,0)$ multiplet provides a spectacular display of the validity of the \emph{`susy without b.c.'} formalism. (General 2D $N=(1,0)$ multiplets with external Lorentz indices, which are required in our construction, are identified in the appendices \ref{unusualMult} and \ref{RedSplitting}.) In section \ref{secMultBC}, we will demonstrate that the b.c. in our model relate the ECTM to the boundary SCM. We then summarize our results, and discuss their extension to higher-dimensional models.

\section{Supersymmetric bulk + boundary system} \label{3DHetM}
\subsection{3D supergravity on a manifold with a boundary}

Our starting point is three-dimensional $N=1$ supergravity in the presence of a boundary as has been considered in \cite{Belyaev:2007bg}. We will follow the same conventions~\footnote{
$M$, $N$ are curved 3D indices, 
$A$, $B$ are flat 3D indices, with decomposition
$M=(m,3)$ and $A=(a,\hat3)$. The 3D gamma matrices satisfy
$\ga^A\ga^B=\ga^{AB}+\eta^{AB}$
with $\eta^{AB}=(-++)$ and $\ga^A\ga^B\ga^C=\ga^{ABC}+
\eta^{AB}\ga^C+\eta^{BC}\ga^A-\eta^{AC}\ga^B$
with $\ga^{ABC}=\eps^{ABC}$. Our spinors are Majorana;
$\psibar=\psi^{\rm T}C$, $C^{\rm T}=-C$, 
$C\ga^A C^{-1}=-(\ga^A)^{\rm T}$. The 3D epsilon tensor is related to the 2D epsilon tensor by $\vare^{ab\hat3} = \vare^{ab}$. } 
and briefly review the results of \cite{Belyaev:2007bg} here. We consider a three-dimensional manifold with a single boundary normal to the $x^3$ direction, where the bulk runs over the range $0<x^3$. The presence of the boundary breaks the symmetry under translations in the $x^3$ direction and, as the susy algebra closes on these translations, half the supersymmetry is broken as well. In the conventions used, the surviving supersymmetry is parametrized by $\e_+$, which is related to the bulk supersymmetry parameter $\e=\e_{+}+\e_{-}$ by $\e_+ = P_+ \e$ with the projection operators defined by $P_\pm = \fr12 (1 \pm \g^3 )$. Much of the algebra, in this bulk + boundary set up, is simplified by the unusual Lorentz gauge choice,
\be
\label{gauge}
e_a{}^3=0 \qrq e_m{}^{\hat3}=0
\ee
(whereas $e_3{}^a\neq 0$ and $e_{\hat3}{}^m\neq0$).
This condition is not invariant under either the bulk susy $(\d_Q)$ or Lorentz $(\d_L)$ transformations. However, \eqref{gauge} is invariant under the modified supersymmetry transformation,
\bea
\label{msusy}
\da_Q^\rr(\ep_{+})=\da_Q(\ep_{+})
+\da_L(\la_{a\hat3}=-\epbar_{+}\psi_{a-}).
\eea
This modified susy represents the supersymmetry transformations intrinsic to the boundary and involves a standard supersymmetry transformation, combined with a compensating Lorentz transformation, which restores the gauge choice for the boundary vielbein. In what follows, we will see that fields, which transform under $\d_Q$ and $\d_L$ in the bulk, can be formed into well-behaved multiplets transforming under $\d^\pr_Q$ on the boundary. 

The bulk we consider is populated by a 3D supergravity multiplet $(e_M{}^A,\psi_M,S)$ which transforms under the $\d_\e=\d_Q(\e)$ susy as
\bea
\label{susy1}
\da_\ep e_M{}^A=\epbar\ga^A\psi_M, \quad
\da_\ep\psi_M=2\wh D_M\ep, \quad
\da_\ep S=\half\epbar\ga^{MN}\wh\psi_{MN} ,
\eea
where
\ba 
\wh D_M\ep &= D_M(\wh\om)\ep+\frac{1}{4}\ga_M\ep S, &
\wh {\p}_{MN} &= \wh D_M \p_N -\wh D_N \p_M, \nn \\
\wh D_M\psi_N &= D_M(\wh\om)\psi_N-\frac{1}{4}\ga_N\psi_M S, &
D_M(\wh\om)\psi_N &=\pa_M\psi_N+\frac{1}{4}\wh\om_{MAB}\ga^{AB}\psi_N .
\label{DefDeriv}
\ea
Here $\wh D_M$ is the 3D-supercovariant derivative.~\footnote{
A supercovariant quantity has supersymmetry variation which does not involve derivatives of the supersymmetry parameter $\e$. Acting on a supercovariant quantity with the supercovariant derivative produces another supercovariant quantity.
} 
It is covariant under Lorentz transformations but is not covariant under diffeomorphisms. The supercovariant spin connection which appears in this derivative is given by
\ba
\label{spincon}
&\wh\om_{MAB}=\om(e)_{MAB}+\kappa_{MAB}, \quad
\kappa_{MAB}=\frac{1}{4}(\psibar_M\ga_A\psi_B-\psibar_M\ga_B\psi_A
+\psibar_A\ga_M\psi_B), & \nn\\
&\om(e)_{MAB}=\half(C_{MAB}-C_{MBA}-C_{ABM}), \quad
C_{MN}{}^A=\pa_M e_N{}^A-\pa_N e_M{}^A ,&
\ea
and it transforms under supersymmetry as
\be
\label{omMABtr}
\da_\ep\wh\om_{M A B}=\half\epbar(\ga_B\wh\psi_{M A}-\ga_A\wh\psi_{M B}-\ga_M\wh\psi_{A B})
-\half(\epbar\ga_{A B}\psi_M)S .
\ee
The supergravity multiplet also transforms under the bulk Lorentz transformations $\d_\la=\d_L(\la_{AB})$ as
\ba
\da_\la e_M{}^A&=\la^{AB}e_{MB},&
\da_\la\psi_M &=\frac{1}{4}\la^{AB}\ga_{AB}\psi_M, \nn \\
\da_\la S&=0, &
\da_\la\wh\om_{MAB} &=-D(\wh\om)_M\la_{AB} .
\label{Lorentz}
\ea
We define the 3D Riemann tensor $R(\wh\om)_{MN}{}^{AB}=\pa_M\wh\om_N{}^{AB}
+\wh\om_M{}^{AC}\wh\om_{NC}{}^B-(M\leftrightarrow N)$, and find that the corresponding supercovariant tensor is given by
\be
\bs
\wh R_{M N A B} &= \pa_M\wh\om_{N A B}+\wh\om_{M A}{}^C\wh\om_{N C B}
+\frac{1}{8}(\psibar_M\ga_{A B}\psi_N)S  \\
&-\qter\psibar_M(\ga_B\wh\psi_{N A}-\ga_A\wh\psi_{N B}-\ga_N\wh\psi_{A B})-(M\lra N) .
\es
\label{DefHatR}
\ee
Supersymmetry variation of the supercovariant gravitino field strength is then
\be
\da_\ep\wh\psi_{A B}  =\qter\ga^{C D}\ep\wh R_{A B C D}+\half\ga_B\ep\wh D_A S+\frac{1}{8}\ga_{A B}\ep S^2-(A\lra B),
\label{psiABtr}
\ee
where $\wh D_M S=\pa_M S-\qter\psibar_M\ga^{BC}\wh\psi_{BC}$.

With the scalar curvature defined by $R(\wh\om)=e_B{}^M e_A{}^N R(\wh\om)_{MN}{}^{AB}$, the standard 3D $N=1$ supergravity action is 
\bea
\label{FASG}
S_{SG}=\int_\mc{M} d^3x e_3\Big[ \frac{1}{2}R(\wh\om)
+\frac{1}{2}\psibar_M\ga^{MNK}D(\wh\om)_N\psi_K
+\frac{1}{4}S^2 \Big] .
\eea
In usual discussions of supersymmetry, one considers a Lagrangian invariant if it varies into a total derivative. However, in the model considered here, the bulk Lagrangian lives on a manifold $\cM$ that has a boundary $\pa \cM$. This means that when the bulk action is varied, the  total derivative produced is mapped into a surface term on the boundary. The presence of this surface term means that the action is no longer supersymmetric unless certain boundary conditions are imposed which force the surface term to vanish. 

The work of \cite{Belyaev:2007bg} improves on this situation by adding a boundary-localized term to the action. The variation of this boundary term cancels the surface term produced by the variation of the bulk. This gives an action that is supersymmetric, under the modified transformations \eqref{msusy}, \emph{without the need for any boundary conditions.}  This improved supergravity action is given by 
\bea
\label{imprSG}
S_{SG}^\text{impr} &=& \int_\mc{M} d^3x e_3  \bls \frac{1}{2}R(\wh\om)
+\frac{1}{2}\psibar_M\ga^{MNK}D(\wh\om)_N\psi_K
+\frac{1}{4}S^2 \brs \nn\\
&& \quad+\int_{\pa \mc{M}} d^2x e_2 \bls
\wh K+\half\psibar_{a+}\ga^a\ga^b\psi_{b-} \brs,
\eea
where $\wh K=e^{ma}\wh K_{ma}$ and 
$\wh K_{ma}=\wh\om_{ma\hat3}-\half\psibar_{m+}\psi_{a-}$, which is the (symmetric) supercovariant extrinsic curvature.~\footnote{
The extrinsic curvature is usually defined by
$K_{MN}=\pm P_M{}^K P_N{}^L \nabla_K n_L$ where 
$n_M$ is the outward-pointing vector normal to the boundary, $P_M{}^K=\da_M{}^K-n_M n^K$ projects into the directions tangent to the boundary and 
$\nabla_K n_L=\pa_K n_L-\Ga_{KL}{}^S n_S$.
In our gauge and with our choice of coordinates, 
$n_M=(0,0,-e_3{}^{\hat3})$
and
$K_{mn}=\mp\Ga_{mn}{}^3 n_3=\pm\Ga_{mn}{}^3 e_3{}^{\hat3}$.
The vielbein postulate yields $\Ga_{mn}{}^3 e_3{}^{\hat3}
=-\om_{ma}{}^{\hat3} e_n{}^a$. Our sign choice is then 
$K_{MN}=- P_M{}^K P_N{}^L \nabla_K n_L$.
}

\subsection{2D induced supergravity}

The transformations of the 3D supergravity multiplet imply that the induced 2D supergravity multiplet is $( e_m{}^a, \p_{m +} )$. This transforms under the modified supersymmetry $\d_\e^\pr=\d_Q^\pr(\e_{+})$ introduced in (\ref{msusy}) as \cite{Belyaev:2007bg}
\be
\da^\rr_\ep e_m{}^a=\epbar_{+}\ga^a\psi_{m+}, \quad
\da^\rr_\ep\psi_{m+}=2 D^\rr_m(\wh\om^{+}) \ep_{+},
\label{isg}
\ee
where $D^\pr_m $ is the induced boundary covariant derivative,~\footnote{
In our conventions, the prime is universally used to mean ``appropriate for the boundary.'' (We could also write $\wh\om^\pr_{mab}$ instead of $\wh\om^{+}_{mab}$.) The 2D supercovariance is with respect to (\ref{msusy}).
}
\be
D^\rr_m(\wh\om^{+})\ep=\pa_m\ep+\frac{1}{4}\wh\om_{mab}^{+}\ga^{ab}\ep,
\ee
and the 2D-supercovariant spin connection $\wh\om_{mab}^{+}$ is defined by
\be
\bs
\wh\om_{mab} &= \wh\om_{mab}^{+}+\kappa_{mab}^{-}, \hspace{100pt} \wh\om_{mab}^{+} = \om(e)_{mab}+\kappa_{mab}^{+},\\
\om(e)_{mab}&=\half(C_{mab}-C_{mba}-C_{abm}) = -C_{abm} , \quad
C_{mn}{}^a=\pa_m e_n{}^a-\pa_n e_m{}^a,\\
\kappa_{mab}^{-} &= \frac{1}{4}(\psibar_{m-}\ga_a\psi_{b-}
-\psibar_{m-}\ga_b\psi_{a-}+\psibar_{a-}\ga_m\psi_{b-}) = \frac{1}{2}\psibar_{a-}\ga_m\psi_{b-},
\\
\kappa_{mab}^{+} &=\frac{1}{4}(\psibar_{m+}\ga_a\psi_{b+}
-\psibar_{m+}\ga_b\psi_{a+}+\psibar_{a+}\ga_m\psi_{b+}) = \frac{1}{2}\psibar_{a+}\ga_m\psi_{b+}.
\es
\label{def2Dspincon}
\ee
To simplify the expressions for $\kappa_{mab}$ and $\om_{mab}$, we have used the 2D Schouten identity, which states that the antisymmetrization of any 3 indices vanishes. With these definitions, $\wh \o^+_{mab}$ is the standard supercovariant spin connection for the induced vielbein $e_m{}^a$.  The 2D-supercovariant gravitino field strength and Riemann tensor are defined by
\ba
\wh \p^\pr_{mn+} &= D^\pr_m(\wh\om^{+}) \p_{n+} - (m \lra n) , \nn\\
\wh R^\pr(\wh \o^+)_{mn}{}^{ab} &= \pa_m \wh\o^+_n{}^{ab} + \wh \o^+_{m}{}^{ac} \wh \o^+_{nc}{}^{b} 
+ \fr12 \ov\p_{m+} \g_n \wh \p^{\pr a b}_+ - (m \lra n), 
\label{Def2DRPsi}
\ea
and the analogs of (\ref{omMABtr}) and (\ref{psiABtr}) are quite simple,
\bea
\da_\ep^\pr\wh\om_{mab}^{+}=-\epbar_{+}\ga_m\wh\psi_{ab+}^\pr, \quad
\da_\ep^\pr\wh\psi_{ab+}^\pr=\half\ga^{cd}\ep_{+}\wh R_{abcd}^\pr.
\eea
These are all \emph{standard} 2D $N=(1,0)$ supersymmetry results which follow from the fact that the modified supersymmetry transformations (\ref{msusy}) close into \emph{standard} 2D $N=(1,0)$ supersymmetry algebra \cite{Belyaev:2007bg}. As a result, we can use the standard supergravity tensor calculus \cite{Ferrara:1978jt,Ferrara:1978wj,Stelle:1978yr,Stelle:1978wj,Uematsu:1986aa} to construct (separately) supersymmetric boundary actions depending on boundary-localized fields.

\subsection{Boundary-localized matter} \label{BoundaryMatter}

Now we wish to consider coupling additional boundary-localized matter to the system. The virtue of the \emph{`susy without b.c.'} setup is that this is easily done, as the bulk and boundary are separately supersymmetric. 

The basic 2D $N=(1,0)$ multiplets are the scalar multiplet $\F_2(A) = ( A, \c_-) $ and the spinor multiplet $\P_2(\c_+) = (\c_+, F)$. The fields in these multiplets transform as
\ba
\label{BasicMultiplets}
\dep A &= \ov \e_+  \c_-, & \dep \c_- &= \g^a \e_+ \wh D^\pr_a A, \nn \\
\dep \c_+ &= F\e_+, & \dep F &= \ov \e_+ \g^a \wh D^\pr_a \c_+  ,
\ea
where $\wh D^\pr_m$ is the 2D-supercovariant derivative. It is (minimally) supercovariant with respect to the 2D (induced) supersymmetry $\dep$ and is given by
\ba
\wh D_m^\pr A&=\pa_m A-\half\psibar_{m+}\chi_{-} , &
\wh D_m^\pr\chi_{+}&=D_m^\pr(\wh\om^{+})\chi_{+}-\half F\psi_{m+} .
\ea
According to the 2D $N=(1,0)$ tensor calculus \cite{Uematsu:1986aa}, the multiplets can be multiplied
\ba
\F_2(A) \times \F_2(\tilde A) &= (A \tilde A,\ \tilde A \c_- + A \tilde\c_-) \equiv \F_2( A \tilde A), \nn\\
\P_2(\c_+) \times \F_2(A) &= ( \c_+ A,\ F A - \ov \c_- \c_+ ) \equiv \P_2(\c_+ A) ;
\ea
their derivatives exist in the form of kinetic multiplets,
\ba
T\F_2(A)  &=  ( \g^a \wh D^\pr_a \c_-,\ \wh D^{\prime a} \wh D^\pr_a A ) \equiv \P_2( \g^a \wh D^\pr_a \c_- ),\nn\\
T\P_2 (\c_+)  &=  ( F,\ \g^a \wh D^\pr_a \c_+) \equiv \F_2( F ) ;
\ea
and functions of the scalar multiplet can be defined,
\bea
\F_2( U(A)) = ( U(A),\ U^\pr (A) \c_-),
\eea
where $U^\pr(A) \equiv \pa U(A)/\pa A $. Finally, locally supersymmetric actions are constructed from spinor multiplets as
\be
\int d^2x e_2 \bls F + \fr12 \ov \p_{a+} \g^a \c_+ \brs.
\label{FAct}
\ee

The boundary action we will consider consists of three separately supersymmetric terms. Firstly, there are the kinetic terms for the scalar multiplet formed from the multiplet $\F_2(A) \times T \F_2(A)$; similarly, there are the kinetic terms for the spinor multiplet formed from the multiplet  $\P_2(\c_-) \times T \P_2(\c_-)$; and finally, there are superpotential-type terms formed from the multiplet $\P_2(\c_-) \times \F_2( U(A) )$. This gives the boundary matter action
\be
\label{BoundaryAct}
\bs
S_m =  a &\int_{\pa M}  d^2 x e_2 \bls - \pa_a A \pa^a  A  - \ov \c_- \g^a
\pa_a \c_- + \ov \p_{a+} \g^b \g^a \c_-\pa_b A \brs \\
 + \ b &\int_{\pa M} d^2 x e_2 \bls F^2 - \ov \chi_+ \g^a \pa_a \c_+ \brs\\
 + \ c &\int_{\pa M} d^2 x e_2 \bls U(A) F - \ov \c_+ \c_- U^\pr(A) + \fr12 \ov \p_{a+} \g^a \c_+ U(A) \brs,
\es
\ee
where $a$, $b$ and $c$ are constants put in to keep track of the contributions arising from these three separately supersymmetric terms. The complete action, $S=S_{SG}^\text{impr}+S_m$, formed from \eqref{imprSG} + \eqref{BoundaryAct}, is supersymmetric as each of its parts are separately supersymmetric. This shows how easy it is to create bulk + boundary actions similar to those in \cite{Bagger:2002rw,Moss:2004ck,Bagger:2004rr,Falkowski:2005fm,Belyaev:2005rt,Pugh:2010ii} when the \emph{`supersymmetry without b.c.'} formalism is employed.~\footnote{
We note that the \emph{`susy without b.c.'} formalism has recently been used in \cite{Berman:2009kj,Grumiller:2009dx}.
}

\subsection{Field equations and boundary conditions} \label{BoundaryConditions}

When the variational principle is applied to the complete action, three classes of equation arise from the three quite different sectors of the variational principle. The variation of bulk fields in the bulk gives rise to field equations for these bulk fields in the standard way. These bulk field equations can be stated as the following \emph{supercovariant} equations, 
\ba
\wh R_{AB}  - \fr12 \eta_{AB} \wh R &= 0,& \wh \p_{AB} &=0, & S&= 0,
\label{BulkEOM}
\ea
where the supercovariant Ricci tensor is defined by $\wh R_{M B}=e^{N A}\wh R_{M N A B}$.
These vary into one another under the bulk supersymmetry and under the bulk Lorentz transformations, and hence vary into each other under the induced supersymmetry as well. We describe this property by saying that this set of equations forms an orbit.

Similarly, the variation of boundary fields on the boundary gives rise to a second set of field equations for the boundary fields. These boundary field equations can also be stated in the supercovariant form,
\ba
&2 a  \wh D^\pr_a \wh D^{\pr a} A + c  F U^\pr(A) - c \ov \c_+ \c_- U\ppr(A)  = 0, &
2b F + c  U(A) &=0, \nn \\
&2 a  \g^a \wh D^\pr_a \c_-+ c  \c_+ U^\pr(A) = 0 ,&
2b  \g^a \wh D^\pr_a \c_+ + c  U^\pr(A) \c_- &=0.
\label{BoundaryEOM}
\ea
As with the bulk field equations, this set forms another orbit under the induced supersymmetry transformations. 

Finally, there are the equations implied by the variation of bulk fields on the boundary. This includes terms which are present due to having used integration by parts when deriving the bulk field equations \eqref{BulkEOM} as well as terms which arise due to the variation of the boundary localized terms in \eqref{imprSG}. This gives the boundary conditions, which once again can be stated in the supercovariant form,~\footnote{
We consider only Neumann (natural) boundary conditions which arise from \emph{unrestricted} variations of $e_m{}^a$ and $\psi_{m+}$ on the boundary. The other possibility is to use Dirichlet boundary conditions which restrict the variations of these fields on the boundary.
} 
\ba
\wh K_{ab} - \eta_{ab} \wh K \bc & = a \bls 2 \inD_a A \inD_b A +  \ov \c_- \g_b \inD_a \c_-  - \h_{a b}  \inD_c A \wh D^{\pr c}  A  - \h_{a b} \ov \c_- \g^c
\inD_c \c_- \brs \nn \\
&
+ b \bls \ov \c_+ \g_b \inD_a \c_+ + \h_{a b}   F^2 - \h_{a b} \ov \chi_+ \g^c \inD_c \c_+ \brs \nn \\
& + c \bls \h_{a b}     U(A) F - \h_{a b} \ov \c_+ \c_- U^\pr(A) \brs,  \nn \\
\p_{a-} \bc&= a \g^b \g_a \c_- \wh D^\pr_b A - \fr{c}{2} \g_a \c_+ U(A). 
\label{KPsiBC}
\ea
These two equations do not form an orbit on their own. However, the fact that (\ref{BulkEOM}), (\ref{BoundaryEOM}) and (\ref{KPsiBC}) have all been derived from the supersymmetric action via the variational principle guarantees that \emph{together} they are closed under supersymmetry variation. The question is how to extract a \emph{minimal} orbit that contains (\ref{KPsiBC}) and some of the bulk and/or boundary field equations. Furthermore, this orbit should be expressible as a multiplet of the induced supersymmetry.~\footnote{
If one could formulate the bulk (3D) supergravity in terms of boundary (2D) superfields, then the field equations and boundary conditions would automatically arise in the superfield form. The way this argument lifts to the component (tensor calculus) analysis was discussed in the rigidly supersymmetric setting in \cite{Belyaev:2005rs,Belyaev:2008xk}.
} 
In what follows, we will identify what this multiplet is and discuss its physical origin.

\subsection{ Supersymmetry with/without boundary conditions} 

In the \emph{`susy without b.c.'} formalism we have described, the three classes of field equation encountered in section \ref{BoundaryConditions} appear on similar footings. However, our motivation for considering this 3D system is to gain a fuller understanding of more general and physically important bulk + boundary systems where such a formalism is not always possible. This is because the \emph{`susy without b.c.'} formalism relies heavily on the existence of auxiliary fields. When these auxiliary fields are not available, the best one can do, in general, is to construct a bulk + boundary system where supersymmetry of the action relies on using the boundary conditions. In the resulting \emph{`susy with b.c.'} formalism, the boundary conditions are thus set apart from the other equations implied by the variational principle. As studies in 5D \cite{Altendorfer:2000rr,Bagger:2002rw,Bagger:2004rr,Falkowski:2005fm,Belyaev:2005rt}, 7D \cite{Pugh:2010ii} and 11D \cite{Horava:1996ma,Moss:2003bk,Moss:2004ck} have demonstrated, it is quite non-trivial to achieve consistency within the \emph{`susy with b.c.'} formalism. It is for this reason that we wish to obtain a fuller understanding of the bulk + boundary systems in the simpler setups where both formulations can be used.

\section{The Extrinsic Curvature Tensor Multiplet} \label{secECTM}

The first step in enabling the boundary conditions we have identified to be phrased as a multiplet of the induced supersymmetry, is finding a multiplet which contains the extrinsic curvature tensor $\wh K_{ab}$. We begin this process by noting that the variation of the odd parity gravitino is given by \cite{Belyaev:2007bg}
\bea
\da_\ep^\pr\psi_{a-}=\ga^b\ep_{+}(\wh K_{a b}+\half \eta_{a b}S). 
\label{varPsi-}
\eea
Using that $\eps^{a b}\equiv \eps^{a b\hat3}$, $\ga^{a b}=\eps^{a b\hat3}\ga^{\hat3}$ and $\ga^a=\ga^{a b}\ga_b=-\eps^{a b}\ga_b\ga^{\hat3}$ imply
\bea
\ga^a\ep_{+}=-\eps^{a b}\ga_b\ep_{+},
\eea
we rewrite \eqref{varPsi-} in the alternative form,
\bea
\label{Uab}
\boxed{
\da_\ep^\pr\psi_{a-}=\ga^b\ep_{+}U_{b a}, \quad
U_{a b}\equiv \wh K_{a b}+\half\eps_{a b}S.
}
\eea
Given that $\wh K_{a b}$ is symmetric, whereas $\eps_{a b}$ is antisymmetric, we see that in this second form $S$ enters independently, without mixing with $\wh K_{a b}$. The general 2D $N=(1,0)$ multiplet that contains the transformation (\ref{Uab}) is identified in appendix \ref{unusualMult}. It is a \emph{reducible} multiplet $(\zeta_{a-},U_{a b},\la_{a+})$ with the transformations given in \eqref{Ma-}. In this section, we will establish that the complete extrinsic curvature tensor multiplet (ECTM) is given by
\be
\boxed{
\text{ECTM} = \Big(\psi_{a-},\quad \wh K_{a b}+\half\eps_{a b}S, \quad -\wh\psi_{a\hat3+}\Big).
}
\label{ECTM}
\ee 

\subsection{The variation of the middle component}

First, we will demonstrate that  the middle component,  $U_{a b}=\wh K_{a b}+\half\eps_{a b}S$ with 
$\wh K_{ma}=\wh\om_{ma\hat3}-\half\psibar_{m+}\psi_{a-}$, transforms in the correct way. Using the unmodified supersymmetry (\ref{omMABtr}) and Lorentz (\ref{Lorentz}) transformations of $\wh\om_{M A B}$, we find that the modified (or induced) supersymmetry transformation (\ref{msusy}) of $\wh\om_{m a\hat3}$ is
\be
\da_\ep^\pr\wh\om_{m a\hat3} = -\half\ep_{+}(
\wh\psi_{m a-}+\ga_a\wh\psi_{m\hat3+}+\ga_m\wh\psi_{a\hat3+})
-\half(\epbar_{+}\ga_{a}\psi_{m+})S+D(\wh\om)_m(\epbar_{+}\psi_{a-}).
\ee
Analyzing the covariant derivative which appears in this equation, we note that
\bea
D(\wh\om)_m(\epbar_{+}\psi_{a-})
&=& \pa_m(\epbar_{+}\psi_{a-})+\wh\om_{m a}{}^b(\epbar_{+}\psi_{b-}) \nn\\
&=& D_m^\pr(\wh\om^{+})(\epbar_{+}\psi_{a-})+\half(\psibar_{a-}\ga_m\psi_{b-})(\epbar_{+}\psi_{-}^b), 
\eea
and using the Fierz identity,~\footnote{
The 2D Fierz identities read
$(\epbar_{+}\psi_{-})\eta_{+}=
-\half(\epbar_{+}\ga^c\eta_{+})\ga_c\psi_{-}$
and
$(\epbar_{+}\psi_{-})\phi_{-}
=-(\epbar_{+}\phi_{-})\psi_{-}$.
}
we find that the last term vanishes because
\bea
\psi_{-}^b(\chibar_{+}\psi_{b-})=-\half\ga_a\chi_{+}(\psibar_{b-}\ga^a\psi_{-}^b)=0 . 
\eea
The variation of the supercovariant extrinsic curvature is then given by
\bea
\da_\ep^\pr\wh K_{m a} &=& \da_\ep^\pr\wh\om_{m a\hat3}-\half\psibar_{m+}\da_\ep^\pr\psi_{a-}
-\half\psibar_{a-}\da_\ep^\pr\psi_{m+} \nn\\
&=& -\half\epbar_{+}(\wh\psi_{m a-}+\ga_a\wh\psi_{m\hat3+}+\ga_m\wh\psi_{a\hat3+})
-\half(\epbar_{+}\ga_a\psi_{m+})S 
+D_m^\pr(\wh\om^{+})(\epbar_{+}\psi_{a-}) \nn\\
&& -\psibar_{a-}D_m^\pr(\wh\om^{+})\ep_{+}
-\half(\psibar_{m+}\ga^b\ep_{+})(\wh K_{a b}+\half\eta_{a b}S) \nn\\
&=& \epbar_{+}D_m^\pr(\wh\om^{+})\psi_{b-}+\half(\epbar_{+}\ga^b\psi_{m+})(\wh K_{a b}-\half\eta_{a b}S) \nn\\
&& -\half\epbar_{+}(\wh\psi_{m a-}+\ga_a\wh\psi_{m\hat3+}+\ga_m\wh\psi_{a\hat3+}).
\label{varK1}
\eea
Flattening the indices with the induced vielbein gives
\bea
\da_\ep^\pr\wh K_{a b} &=& e_a{}^m\da\wh K_{m b}-(\epbar_{+}\ga^c\psi_{a+})\wh K_{c b} \nn\\
&=& \epbar_{+}D_a^\pr(\wh\om^{+})\psi_{b-}-\half(\epbar_{+}\ga^c\psi_{a+})(\wh K_{b c}+\half\eta_{b c}S) \nn\\
&&-\half\epbar_{+}(\wh\psi_{a b-}+\ga_b\wh\psi_{a\hat3+}+\ga_a\wh\psi_{b\hat3+}).
\eea
Noting that the minimally supercovariant derivative of $\psi_{a-}$ is given by
\bea
\wh D_m^\pr\psi_{a-} \equiv D_m^\pr(\wh\om^{+})\psi_{a-}-\half\ga^b\psi_{m+}U_{b a},
\eea
where
$
D_m^\pr(\wh\om^{+})\psi_{a-}=\pa_m\psi_{a-}+\qter\wh\om_{m b c}^{+}\ga^{b c}\psi_{a-}+\wh\om^{+}_{m a}{}^b\psi_{b-}
$,
we can rewrite \eqref{varK1} as
\bea
\boxed{
\da_\ep^\pr\wh K_{a b} = \epbar_{+}\wh D_a^\pr\psi_{b-}
-\half\epbar_{+}(\wh\psi_{a b-}+\ga_b\wh\psi_{a\hat3+}+\ga_a\wh\psi_{b\hat3+}).
}
\label{varK2}
\eea
Let us now analyze $\wh\psi_{a b-}$ which appears in this equation. Starting with
\bea
\wh\psi_{M N}=\pa_M\psi_N+\qter\wh\om_{M A B}\ga^{A B}\psi_N-\qter\ga_N\psi_M S-(M\lra N),
\label{defPsiMN}
\eea
then restricting the indices to lie tangent to the boundary and projecting with the negative chirality projection matrix $P_{-}=\half(1-\ga^{\hat3})$, we find that
\bea
\wh\psi_{m n-}=\pa_m\psi_{n-}+\qter\wh\om_{m a b}\ga^{a b}\psi_{n-}+\half\wh\om_{m a\hat3}\ga^a\psi_{n+}
-\qter\ga_n\psi_{m+}S-(m\lra n).
\eea
From the definition of the induced spin connection \eqref{def2Dspincon}, we have
\bea
\pa_m\psi_{n-}-(m\lra n)=e_n{}^a\pa_m\psi_{a-}+\om(e)_{m n}{}^a\psi_{a-}-(m\lra n).
\eea
Substituting this into \eqref{defPsiMN} gives
\ba
\wh\psi_{m n-} &= e_n{}^a D_m^\pr(\wh\om^{+})\psi_{a-}
+\half(\wh K_{m a}+\half e_{m a}S)\ga^a\psi_{n+} \nn\\
&\hspace{-35pt}
-\half\psi_{b-}(\psibar_{n+}\ga_m\psi_{b+})
+\frac{1}{8}\ga^{b c}\psi_{n-}(\psibar_{b-}\ga_m\psi_{c-})
+\qter\ga^a\psi_{n+}(\psibar_{m+}\psi_{a-})-(m\lra n).
\ea
After some Fierzing, we find that the 3-Fermi terms in the second line vanish. Thus
\bea
\wh\psi_{a b-}= D_a^\pr(\wh\om^{+})\psi_{b-}+\half(\wh K_{a c}+\half\eta_{a c}S)\ga^c\psi_{b+}-(a\lra b),
\eea
and therefore
\bea
\boxed{
\wh\psi_{a b-}=\wh D_a^\pr\psi_{b-}-(a\lra b).
}
\eea
Substituting this back into \eqref{varK2}, we find that the variation of $\wh K_{a b}$  becomes manifestly $(a\lra b)$ symmetric,
\bea
\boxed{
\da_\ep^\pr\wh K_{a b} = \half\epbar_{+}\Big(
\wh D_a^\pr\psi_{b-}-\ga_a\wh\psi_{b\hat3+}+(a\lra b)\Big).
}
\eea
Next, we note that the variation of the auxiliary field $S$ can be written as
\bea
\da_\ep^\pr S &=& \half\epbar_{+}\ga^{a b}\wh\psi_{a b-}+\epbar_{+}\ga^a\wh\psi_{a\hat3+} \nn\\
&=& -\eps^{a b}\Big(\half\epbar_{+}\wh\psi_{a b-}+\epbar_{+}\ga_b\wh\psi_{a\hat3+}\Big).
\eea
Now using the identity $\eps_{a b}\eps^{c d}=-(\da_a^c\da_b^d-\da_a^d\da_b^c)$ yields
\bea
\eps_{a b}\da_\ep^\pr S &=& \epbar_{+}\Big\{\wh\psi_{a b-}-\Big[\ga_a\wh\psi_{b\hat3+}-(a\lra b)\Big]\Big\}  \nn\\
&=& \epbar_{+}\Big(\wh D_a^\pr\psi_{b-}-\ga_a\wh\psi_{b\hat3+}-(a\lra b)\Big),
\eea
and therefore
\bea
\boxed{
\da_\ep^\pr U_{a b}=\epbar_{+}\Big(\wh D_a^\pr\psi_{b-}-\ga_a\wh\psi_{b\hat3+}\Big).
}
\eea
This shows that $U_{a b}$ does indeed transform as the bosonic component of the multiplet \eqref{Ma-}, and identifies $-\wh\psi_{a\hat3+}$ as the top component of the ECTM. 

\subsection{The variation of the top component}

The remainder of the proof is to show that $-\wh\psi_{a\hat3+}$ transforms as required. The modified supersymmetry transformation (\ref{msusy}) of $\wh\psi_{A B}$ is
\bea
\da_\ep^\pr\wh\psi_{A B}=\da_\ep\wh\psi_{A B}+\qter\la_{C D}\ga^{C D}\wh\psi_{A B}
+\la_A{}^C\wh\psi_{C B}+\la_B{}^C\wh\psi_{A C},
\eea
where $\da_\ep\wh\psi_{A B}$ is given in (\ref{psiABtr}), and $\la_{a b}=0$, $\la_{a\hat3}=-\epbar_{+}\psi_{a-}$. Restricting one index to lie in the tangent to the boundary direction ($A=a$) and the other in the normal to the boundary direction ($B=\hat3$), and then projecting with the positive chirality projection matrix $P_{+}=\half(1+\ga^{\hat3})$, we find
\bea
\da_\ep\wh\psi_{a\hat3+}=\half\ga^{b c}\ep_{+}\wh R_{b c a\hat3}+\half\ep_{+}\wh D_a S
=\half\ep_{+}\Big(\eps^{b c}\wh R_{b c a\hat3}+\wh D_a S\Big).
\label{varPa3+1}
\eea
Let us now transform this expression further. The bulk-supercovariant derivative of $S$ is related to the boundary-supercovariant derivative by
\be
\wh D_a S = \wh D_a^\pr S-\qter\psibar_{a-}(\ga^{c d}\wh\psi_{c d+}-2\ga^c\wh\psi_{c\hat3-}), 
\label{defDS}
\ee
where the boundary-supercovariant derivative in question is given by
\be
\wh D_a^\pr S = \pa_a S-\qter\psibar_{a+}(\ga^{c d}\wh\psi_{c d-}+2\ga^c\wh\psi_{c\hat3+}).
\ee
The bulk-supercovariant gravitino field strength is related to the boundary-super\-covariant gravitino field strength by
\bea
\wh\psi_{a b+} &=& \wh\psi_{a b+}^\pr+\Big(\half\ga^c\wh K_{b c}+\qter\ga_a\psi_{b-}S-(a\lra b)\Big).
\label{PtoPpr}
\eea
Analyzing the bulk-supercovariant Riemann tensor defined in (\ref{DefHatR}), we find (after some algebra) the following supercovariant Gauss-Codazzi equation
\bea
\boxed{
\bay[b]{rcl}
\wh R_{a b c\hat3} &=&\dst \wh D_a^\pr\wh K_{b c}+\frac{3}{8}\psibar_{c-}\wh\psi_{a b+}^\pr
+\qter\psibar_{a-}(\ga_c\wh\psi_{b\hat3-}+\ga_b\wh\psi_{c\hat3-})
-\frac{3}{32}(\psibar_{a-}\ga_c\psi_{b-})S \\[10pt]
&&\dst
+\half(\psibar_{c-}\ga_a\psi^d_{-})\wh K_{b d}
-\frac{1}{16}(\psibar_{a-}\ga^d\psi_{b-})\wh K_{c d}
-(a\lra b),
\eay
} \quad
\label{SGCeq}
\eea
where
\bea
\wh D_a^\pr\wh K_{b c}=D_a^\pr(\wh\om^{+})\wh K_{b c}
-\qter\psibar_{a+}\Big(\wh D_b^\pr\psi_{c-}-\ga_b\wh\psi_{c\hat3+}+(b\lra c)\Big).
\eea
Finally, substituting \eqref{defDS}, \eqref{PtoPpr} and \eqref{SGCeq} into \eqref{varPa3+1}, gives
\bea
\boxed{
\da_\ep^\pr\wh\psi_{a\hat3+}=-\ga^{c d}\ep_{+}\wh D_d^\pr U_{c a}
+\qter\ep_{+}(\psibar_{a-}\ga^{c d}\wh\psi_{c d +}^\pr)
-\ep_{+}(\psibar_{-}^b\wh\psi_{a b+}^\pr),
}
\eea
as required for consistency with (\ref{Ma-}). This completes the proof that (\ref{ECTM}) transforms as a (reducible) 2D $N=(1,0)$ multiplet under the modified supersymmetry transformations (\ref{msusy}).

\subsection{Irreducible submultiplets of the ECTM}

Applying the splitting of the reducible multiplet described in (\ref{irreps}), we find that
\bea
\label{irredECTM}
 \P_2(\ga^a\psi_{a-}) &=& \Big(\ga^a\psi_{a-}, \quad \wh K+S\Big), \nn\\
 \P_2(\ga_a\ga^c\ga_b\psi_{c-}) &=& \Big(\ga_a\ga^c\ga_b\psi_{c-}, \quad 4P_{+a}{}^c P_{+b}{}^d\wh K_{c d}\Big), \nn\\
 \F_2(\wh K-S) &=& \Big(\wh K-S, \quad -2\ga^a\wh\psi_{a\hat3+}+\ga^a\ga^b\wh D_b^\pr\psi_{a-}\Big), \nn\\
 \F_2(4P_{-a}{}^c P_{-b}{}^d\wh K_{c d}) &=& \Big(4P_{-a}{}^c P_{-b}{}^d\wh K_{c d}, \quad
-2\ga_a\ga^c\ga_b\wh\psi_{c\hat3+}+\ga_a\ga^c\ga_b\ga^d\wh D_c^\pr\psi_{d-}\Big) \quad
\eea
are the four irreducible submultiplets inside (\ref{ECTM}). The first submultiplet has been identified in \cite{Belyaev:2007bg}, where it was called the `extrinsic curvature multiplet.'

Before closing this section, let us see what happens if one identifies $\wh K_{ab} + \fr12 \h_{ab} S$, entering in \eqref{varPsi-}, with the second component of the multiplet (\ref{Ma-}). This is, in fact, consistent and leads to the following `alternative ECTM' multiplet
\be
\text{altECTM}=
\Big( \p_{a -} \ , \ \wh K_{ab} + \fr12 \h_{ab} S \ , \  \fr14 \g_{a} \g^{c d} \wh \p_{cd-} - \fr12 \g^b \g_a \wh \p_{b \hat3 +} \Big) .
\label{altECTM}
\ee
Subtracting (\ref{ECTM}) from (\ref{altECTM}) yields
\bea
\label{deltaECTM}
\Big(0, \quad P_{- a b}S, \quad \qter\ga_a P_{-}\ga^{C D}\wh\psi_{C D}\Big).
\eea
This is also a multiplet of the (\ref{Ma-}) type, with only a single irreducible submultiplet being non-zero: the $\Phi_2(2 P_{+}^{a b}U_{a b})$ in (\ref{irreps}). This multiplet is set to zero by the bulk field equations (\ref{BulkEOM}), so that the two off-shell multiplets, (\ref{ECTM}) and (\ref{altECTM}), match on-shell.

The above discussion clearly shows that the lowest component of a reducible multiplet does not uniquely determine the other components, whereas in an irreducible multiplet it does. It also makes it clear that the choice of the ECTM is not unique. We prefer the one in (\ref{ECTM}) simply because it is the \emph{minimal} choice.

\section{The supermultiplet of boundary conditions} \label{secMultBC}

Having identified the ECTM, we will now rewrite the boundary conditions we found in section \ref{BoundaryConditions} as a boundary condition on this multiplet. We begin by considering the irreducible submultiplets of the ECTM. These are easier to work with than the reducible multiplet, since for these irreducible multiplets the lowest component uniquely determines the whole multiplet. We construct the multiplets of boundary conditions by substituting \eqref{BoundaryEOM} and \eqref{KPsiBC} into \eqref{irredECTM}, in such a way that the \emph{on-shell}  b.c. \eqref{KPsiBC}, obtained from the variational principle, are lifted to give the  following \emph{off-shell} b.c. 
\ba
\P_2(\ga^a\psi_{a-}) \bc 
&= - c \Big( \c_+ U(A) , \quad F U(A) - \ov \c_+ \c_- U^\pr(A)  \Big), \nn\\
\P_2(\ga_a\ga^c\ga_b\psi_{c-}) \bc
&= 2 a \Big( \g_a \g^c \g_b \c_- A_c , \quad 
4 P_{+a}{}^{c} A_c P_{+b}{}^d A_d + \ov \c_- \g_a \g^c \g_b \chi_{c-} \Big) ,
\nn\\
 \F_2(\wh K-S) \bc
 &= - c \Big( F U(A) - \ov \c_+ \c_- U^\pr(A), \quad \g^a \inD_a \left(  \c_+ U(A) \right) \Big) , \nn\\
\F_2(4P_{-a}{}^c P_{-b}{}^d\wh K_{c d}) \bc
&=  2 a \Big(4 P_{-a}{}^{c} A_c  P_{-b}{}^d A_d  ,\quad 2 \g_a \g^c \g_b \g^d \chi_{c-} A_d  \Big) , \nn\\
&+ 2 b \Big( \ov \c_+ \g_a \g^c \g_b \c_{c+},\quad \g_a \g^c \g_b \c_{c+} F - \g_a\g^c \g_b  \c_+ G_c \Big) ,
\ea
where $A_a=\inD_a A$, $\chi_a=\inD_a\chi$, and $G_a$ is defined in (\ref{LGBH}). These boundary conditions for the submultiplets of the ECTM recombine into the following boundary condition for the ECTM itself~\footnote{
We emphasize that the multiplet on the R.H.S. of (\ref{EOMmult}) is an \emph{off-shell} multiplet. As a boundary condition, (\ref{EOMmult}) reduces to (\ref{KPsiBC}) when the boundary field equations (\ref{BoundaryEOM}) are used.
}
\be
\bs
& \Big( \p_{a-},\quad K_{ab} + \fr12 \vare_{ab} S,\quad - \wh \p_{a \hat3 +} \Big) \bc \\
&= a \Big( \g^b \g_a \c_- \inD_b A ,\quad 
2\inD_a A  \inD_b A  - \eta_{a b} \inD_c A  \wh D^{\pr c} A + \fr12 \ov \c_- \g_a \g^c \g_b \inD_c \c_{-}, \\
& \hspace{150pt} \g^c \g_a \g^d \inD_c \c_{-} \inD_d A  - \fr12 \g_a \g^b \g^c \inD_{c} ( \c_- \inD_b A ) \Big) \\
&+  \fr12 b \Big( 0,\quad \ov \c_+ \g_a \g^c \g_b \inD_c \c_{+} , \quad 
\g^b \g_a \inD_b \c_{+} F - \g^b \g_a \c_+ G_{b} \Big) \\
&-\fr12 c \Big( \g_a \c_+ U(A),\quad 
\eta_{a b} U(A) F - \eta_{ab} \ov \c_+ \c_- U^\pr(A) ,\quad 
\g_{ab} \wh D^{\pr b} ( U(A) \c_+ ) \Big)    .
\label{EOMmult}
\es
\ee
In order to gain some physical insight into this equation, we consider the flat rigidly supersymmetric 2D version of the boundary action (\ref{BoundaryAct}) given by
\be
\bs
S_m^\text{flat} =  a &\int d^2 x  \bls - \pa_a A \pa^a  A  - \ov \c_- \g^a
\pa_a \c_- \brs \\
 + \ b &\int d^2 x  \bls F^2 - \ov \chi_+ \g^a \pa_a \c_+ \brs\\
 + \ c &\int d^2 x  \bls U(A) F - \ov \c_+ \c_- U^\pr(A) \brs .
\es
\ee
The Noether current associated with the invariance  of this action under supersymmetry is the supercurrent 
\be
J_{a -}^\text{flat} = 2 a \g^b \g_a \c_- \pa_b A + c \g_a \c_+ U(A) ,
\ee
whereas the Noether current associated with the invariance under translations is the energy-momentum tensor 
\be
\bs
T_{a b}^\text{flat} & = a \bls 2 \pa_a A \pa_b A +  \ov \c_- \g_b \pa_a \c_-  
- \h_{a b}  \pa_c A \pa^c  A  - \h_{a b} \ov \c_- \g^c\pa_c \c_- \brs \\
&
+ b \bls \ov \c_+ \g_b \pa_a \c_+ + \h_{a b}   F^2 - \h_{a b} \ov \chi_+ \g^c \pa_c \c_+ \brs\\
& + c \bls \h_{a b}     U(A) F - \h_{a b} \ov \c_+ \c_- U^\pr(A) \brs .
\es
\ee
With this in mind, let us return to the locally supersymmetric boundary setup and promote these currents to their (boundary-)supercovariant equivalents, 
\be
\wh J_{a -} = 2 a \g^b \g_a \c_- \wh D^\pr_b A + c \g_a \c_+ U(A)
\ee
and
\be
\bs
\wh{T}_{a b} & = a \bls 2 \inD_a A \inD_b A +  \ov \c_- \g_b \inD_a \c_-  - \h_{a b}  \inD_c A \wh D^{\pr c}  A  - \h_{a b} \ov \c_- \g^c
\inD_c \c_- \brs \\
&
+ b \bls \ov \c_+ \g_b \inD_a \c_+ + \h_{a b}   F^2 - \h_{a b} \ov \chi_+ \g^c \inD_c \c_+ \brs\\
& + c \bls \h_{a b}     U(A) F - \h_{a b} \ov \c_+ \c_- U^\pr(A) \brs. 
\es
\ee
Next, we fit the boundary field equations (\ref{BoundaryEOM}) into two multiplets,~\footnote{
$E_{(\mc{F})}$ denotes the equation of motion obtained through varying the field $\mc{F}$.
} 
\ba
\Big( E_{( \c_-)}, E_{(A)} \Big)  &\equiv 2 a \Big( \g^a  \inD_a \c_-, \inD_a \wh D^{\pr a} A \Big) + c \Big( \c_+ U^\pr(A) , F U^\pr(A) - \ov \c_+ \c_- U\ppr(A) \Big) , \nn \\
\Big( E_{(F)}, E_{( \c_+ )} \Big) &\equiv 2b \Big( F, \ \g^a \inD_a \c_+ \Big) + c \Big( U(A), U^\pr(A) \c_-\Big),
\ea
so that (\ref{BoundaryEOM}) is equivalent to the vanishing of these multiplets.
The Noether currents we have identified can now be combined into a supercurrent multiplet (SCM) \cite{Ferrara:1974pz,Shizuya:2003vm} of the (\ref{Ma-}) type given by
\be
\label{SCM}
\bs
\text{SCM}=
\Big(  \fr12 \wh J_{a - }, &\quad
\wh T_{a b} + \fr14 \ov \c_+ \g_a \g_b  E_{( \c_+ )}  + \fr14 \ov \c_- \g_a \g_b E_{( \c_-)} -  \h_{ab} F E_{( F)} , \\
&\hspace{0pt}
-\fr12 \g^c \inD_c \wh J_{a-} + \fr12 \g^b \g_a E_{( \c_-)}  A_b + \fr14 \g_a \g^b E_{( \c_-)} A_b  \\
&\hspace{0pt}
+\fr14 \g_a \c_- E_{( A)} - \fr14 \g_a E_{(\c_+ )} F - \fr14 \g^b \g_a \c_+ \inD_b E_{(F)} + \fr12\c_{a+} E_{( F)}\Big)
. 
\es
\ee 
Applying the `star transformation' (\ref{star}) to this multiplet gives  
\be
\label{starSCM}
\bs
\Big( - \fr12 \g_{a b} \wh J^b_- , &\quad
\wh T_{b a} - \eta_{a b} \wh T_{c d} \eta^{c d} - \fr14 \ov \c_+ \g_a \g_b E_{( \c_+ )} 
- \fr14 \ov   \c_- \g_a \g_b E_{( \c_- )} + \h_{ab} F E_{(F)} , \\
& \hspace{32pt} 
\fr12 \g^b \g_a E_{(\c_-)} A_b - \fr14 \g_a \g^b E_{(\c_-)} A_b 
- \fr14 \g_a \c_- E_{( A)} \\
& \hspace{30pt}
+ \fr14 \g_a E_{( \c_+ )} F 
- \fr14 \g^b \g_a \c_+ \inD_b E_{(F)}  - \fr12 \g_{a b} \c_+^b  E_{(F)} \Big).
\es
\ee
With a little algebra, we find this to be equal to the R.H.S. of \eqref{EOMmult}. Hence, we conclude that the boundary conditions following from the variational principle can be stated in the following manifestly supersymmetric form~\footnote{
In any number of dimensions, the bulk gravitino kinetic term is $\sim\psibar_M\ga^{M N K}\pa_N\psi_K$, whereas the boundary coupling is $\sim\psi_{m+}J^m$. The boundary condition on the odd parity gravitino is then $\ga^{m n}\psi_{n-}\bc\sim J^m$ \cite{Moss:2004ck,Belyaev:2007aj}. With the ECTM and SCM containing $\psi_{m-}$ and $J_m$, respectively, the `star conjugation' is needed to absorb the $\ga^{m n}$ in the boundary condition.
}
\be
\label{SMBC}
\text{ECTM}\bc=\star\text{SCM},
\ee
where ECTM and $\star$SCM are \emph{off-shell} multiplets given in (\ref{ECTM}) and (\ref{starSCM}), respectively.
This can be interpreted as the supermultiplet equivalent of the Israel junction condition \cite{Israel:1966rt}.

\section{Conclusions}\label{Conclusions}

As we have seen, the boundary conditions in our `3D Heterotic M-Theory' setup can be neatly expressed in the form $\text{ECTM} \bc =\star \text{SCM}$. Here both the Extrinsic Curvature Tensor Multiplet and the Super Current Multiplet are \emph{off-shell} multiplets, thanks to their dependence on auxiliary fields $S$ and $F$. In order to see the implications of our results to higher-dimensional models, where auxiliary fields are not necessarily available, we should discuss what happens when one eliminates these auxiliary fields through their (algebraic) field equations. 

As has been pointed out in \cite{Belyaev:2007bg}, setting $S=0$ in the improved supergravity action (\ref{imprSG}) \emph{preserves} its \emph{`susy without b.c.'} property. This happens because the boundary term in (\ref{imprSG}) does not depend on $S$.~\footnote{
In \cite{Belyaev:2005rs}, it was demonstrated that when one considers boundary actions dependent on bulk auxiliary fields, the elimination of the latter reduces \emph{`susy without b.c.'} to \emph{`susy with b.c.'}
}
As our boundary-localized matter action (\ref{BoundaryAct}) also does not depend on $S$, the on-shell action in our case is also \emph{`susy without b.c.'} Curiously enough, the second submultiplet of the ECTM in (\ref{irredECTM}) is independent of $S$ and thus remains a multiplet in the on-shell case. But for other multiplets, the dependence on $S$ cannot be removed, and so setting $S=0$ necessarily mixes the b.c. multiplets with the bulk field equations. This, however, does not present a conceptual problem because consistency only requires that field equations and boundary conditions \emph{together} form a supersymmetry orbit. And this is always guaranteed if the b.c. are \emph{derived} from the supersymmetric action via the variational principle.

Higher-dimensional supergravity multiplets contain extra fields (scalars, spinors, vectors, antisymmetric tensors) besides the vielbein and the gravitino. The analogs of our ECTM would then include odd parity components of these fields, but the b.c. would still be $\text{ECTM} \bc =\star \text{SCM}$ with either off-shell (if auxiliary fields are available) or on-shell multiplets. For example, in the 5D $N=1$ case, the ECTM would include the odd part of the graviphoton $B_M$ \cite{Belyaev:2005rt}, whereas in the 11D case, the ECTM would include the odd part of the bulk 3-form $C_{M N K}$ \cite{Moss:2004ck}. The b.c. should set these fields equal to conserved currents that are part of the boundary SCM.

We expect that our (off-shell) \emph{`susy without b.c.'} discussion can be repeated in similar 4D and 5D setups. The tensor calculus for 4D $N=1$ supergravity on a manifold with boundary has been constructed in \cite{Belyaev:2008ex}, whereas the more interesting (because of its relation to Randall-Sundrum models) 5D analysis has not yet been performed.~\footnote{
We note that the existing 5D tensor calculus for supergravity on an orbifold \cite{Kugo:2002js} does not satisfy our consistency criteria as there odd parity fields are chosen to vanish at the brane/boundary \cite{Belyaev:2007aj}.
}
We intend to keep working in these directions with the goal of constructing supersymmetric bulk + boundary actions, with the boundary conditions fully compatible with the variational principle.

\vspace{20pt}
\noindent{\bf\large Acknowledgements}
\vspace{10pt}

\noindent
DVB thanks Peter van Nieuwenhuizen for collaboration on early stages of this project. The research of DVB was supported in part by the Department of Energy Grant No. DE-FG02-97ER41029. TGP would like to thank Kelly Stelle for many useful discussions. The work of TGP was supported in part by the STFC.

\vspace{10pt}

\appendix
\section{Multiplets of 2D $N=(1,0)$ supergravity} \label{unusualMult}

The basic irreducible multiplets in the 2D $N=(1,0)$ tensor calculus are the scalar, $\Phi_2(A)=(A,\chi_-)$, and the spinor, $\Psi_2(\chi_+)=(\chi_+, F)$, multiplets whose supersymmetry transformations are given in (\ref{BasicMultiplets}). Besides these, there is a variety of irreducible (and reducible) multiplets with external Lorentz indices \cite{Kugo:1983mv}, which can be found, for example, by applying supercovariant derivatives to the components of the basic multiplets. The variation of these supercovariant objects is given by~\footnote{
In proving these statements, we have used the following useful lemmas describing the commutators of supercovariant derivatives,
\ba
[\inD_a,\inD_b]A &= -\half\chibar_{-}\wh\psi^\pr_{a b+} ,\nn\\
[\inD_a,\inD_b] \inD_c A &= \wh R^\pr_{a b c d} \wh D^{\pr d} A-\half \inD_{c} \ov \c_- \wh\psi^\pr_{a b+}, \nn\\
[\inD_a,\inD_b] \inD_{c} \chi_{-} &= \wh R^\pr_{a b c d} \wh D^{\pr d} \chi_{-}+\qter\wh R^\pr_{a b p q}\ga^{p q} \inD_c \chi_{-} 
-\half\ga^d\wh\psi^\pr_{a b+}\inD_d \inD_c A, \nn\\
[\inD_a,\inD_b]\chi_{+} &= \qter\wh R^\pr_{a b c d}\ga^{c d}\chi_{+}-\half F\wh\psi^\pr_{a b+}, \nn\\
[\inD_a,\inD_b] \inD_c \chi_{+} &= \wh R^\pr_{a b c d} \wh D^{\pr d} \chi_{+} + \qter\wh R^\pr_{a b p q}\ga^{p q} \inD_c \chi_{+}
-\half\wh\psi^\pr_{a b+}G_c. \nn
\ea
Furthermore we note that $ \dep \wh \p_{ab+}^{\pr} = \fr12 \g^{cd} \e_+ \wh R_{abcd}^\pr$
and $\dep\wh R_{a b c d}^\pr=\epbar_{+}\ga_a\inD_b\wh\psi_{c d+}^\pr-(a\lra b)$.
} 
\ba
\dep (\wh D^\pr_a A) &= \epbar_{+} \wh D^\pr_a \chi_{-}, 
& \dep ( \wh D^\pr_a \wh D^\pr_b A )&= \epbar_{+} \wh D^\pr_a \wh D^\pr_b \chi_{-} + \epbar_{+}\ga_a\la_{b+} ,\nn \\
\dep (\wh D^\pr_a \chi_{-} )&= \ga^b\ep_{+} \wh D^\pr_b \wh D^\pr_a A, 
&  \dep( \wh D^\pr_a \wh D^\pr_b \chi_{-}) &= \ga^c\ep_{+} \wh D^\pr_a \wh D^\pr_c \wh D^\pr_b A -\ga_a\ep_{+}B_b ,\nn\\
\dep (\wh D^\pr_a \chi_{+}) &=\ep_{+}G_a,
& \dep (\wh D^\pr_a \wh D^\pr_b \chi_{+} ) &=\ep_{+}H_{a b} .
\ea
where we defined
\ba
\label{LGBH}
\la_{a+} &\equiv -\wh\psi^\pr _{a b+}\inD_b A, \nn\\
G_a &\equiv \wh D^\pr_a F +\half\chibar_{+}\ga^b\wh\psi^\pr_{a b+}, \nn\\
B_a &\equiv \wh D^{\pr b} \chibar_{-}\wh\psi^\pr_{a b+}-\qter \inD_a \chibar_{-}\ga^{c d}\wh\psi^\pr_{c d+}, \nn\\
H_{a b} &\equiv \wh D^\pr_a G_{ b} + \wh D^{\pr c} \chibar_{+}\ga_a\wh\psi^\pr_{b c+}+\half \inD_b \chibar_{+}\ga^c\wh\psi^\pr_{a c+} .
\ea
These new quantities transform as
\ba
\dep\la_{a+}  &= \ga^{c d}\ep_{+}\wh D^\pr_d \wh D^\pr_c \wh D^\pr_a A +\ep_{+}B_a , \nn\\ 
\dep G_a &=\epbar_{+}\ga^b \wh D^\pr_b \wh D^\pr_a \chi_{+}, \nn\\
\dep B_a &= \epbar_{+}\ga^{c d} \wh D^\pr_d \wh D^\pr_c \wh D^\pr_a \chi_{-}
-\epbar_{+}\ga^c(\wh\psi^\pr_{a b+} \wh D^\pr_c \wh D^{\pr b} A+\wh\psi^\pr_{c b+} \wh D^{\pr b} \wh D^\pr_a A) , \nn\\
\dep H_{a b}& =\epbar_{+}\ga^c \wh D^\pr_c \wh D^\pr_a \wh D^\pr_b \chi_{+}.
\ea

We use these transformations to identify several 2D multiplets. There is a scalar multiplet with a single external Lorentz index $(A_a, \c_{a-})=( \wh D^\pr_a A, \wh D^\pr_a \c_-) $ which transforms as
\be
\label{Scalara-}
\boxed{
\dep A_a = \epbar_{+}\chi_{a-},  \quad \dep \chi_{a-} = \ga^b\ep_{+} \wh D^\pr_b A_{a}.
}
\ee
Similarly, there is a scalar multiplet with two external Lorentz indices $(A_{ab}, \c_{ab-}) = (  \inD_a \inD_b A, \wh D^\pr_a \wh D^\pr_b \chi_{-}+\ga_a\la_{b+})$ which transforms as 
\bea
\label{Mab}
\boxed{
\dep A_{a b}=\epbar_{+}\c_{a b-}, \quad
\dep\c_{a b-}=\ga^c\ep_{+}\wh D^\pr_c A_{a b} .
}
\eea
We also find spinor multiplets with external Lorentz indices. There is a spinor multiplet with one external Lorentz index $(\c_{a+}, F_a)=(\wh D^\pr_a \c_+, G_a )$ which transforms as
\bea
\label{Ma+2}
\boxed{
\dep \chi_{a+}=\ep_{+}F_a, \quad \dep F_a=\epbar_{+}\ga^b\wh D^\pr_b \chi_{a+}.
}
\eea
Similarly, there is a spinor multiplet with two external Lorentz indices $(\c_{ab +}, F_{ab})=(\wh D^\pr_a \wh D^\pr_b \c_+ , H_{ab} )$ which transforms as
\bea
\label{Mab+}
\boxed{
\dep\chi_{a b+}=\ep_{+}F_{a b}, \quad \dep F_{a b}=\epbar_{+}\ga^c\wh D^\pr_c\chi_{a b+} .
}
\eea

Besides the above irreducible multiplets, we also identify certain \emph{reducible} multiplets. One such multiplet is $( \z_-, U_{a}, \l_+ ) =( \g^a \inD_a \c_+,\ G_a,\ -\fr14  \c_+ \wh R^\pr + \fr14 \g^{ab} \wh \p_{ab+}^\pr F )$ which transforms as
\be
\boxed{
\bay[b]{rl}
\dep\z_{-} &=\dst \ga^a\ep_{+}U_{ a} , \\[3pt]
\dep U_{a } &=\dst \epbar_{+}\wh D^\pr_a\z_{-}+\epbar_{+}\ga_a\la_{+} , \\
\dep\la_{+} &=\dst \ga^{ab}\ep_{+}\wh D^\pr_b U_{a }
-\qter\ep_{+}(\ov \z_{-}\ga^{c d}\wh\psi_{c d+}^\pr).
\eay
}
\label{Mred}
\ee
The corresponding multiplet with \, one \, external Lorentz index is \, $(\z_{a-}, U_{ab}, \l_{a+})=$ $ (\inD_a \c_{-},\ \inD_a \inD_b A,\ \l_{a+} )$ which transforms as
\be
\boxed{
\bay[b]{rl}
\dep\z_{a-} &=\dst \ga^b\ep_{+}U_{b a} , \\[3pt]
\dep U_{a b} &=\dst \epbar_{+}\wh D^\pr_a\z_{b-}+\epbar_{+}\ga_a\la_{b+} , \\
\dep\la_{a+} &=\dst \ga^{c d}\ep_{+}\wh D^\pr_d U_{c a}
-\qter\ep_{+}(\ov \z_{a-}\ga^{c d}\wh\psi_{c d+}^\pr)
+\ep_{+}(\ov \z^b_{-}\wh\psi_{a b+}^\pr).
\eay
}
\label{Ma-}
\ee
This reducible multiplet is central for the discussion in the main text. The ECTM (\ref{ECTM}) and the SCM (\ref{SCM}) are two examples of this multiplet. Another example is the 2D $N=(1,0)$ Ricci tensor multiplet,
\bea
(\ga^b\wh\psi^\pr_{a b+}, \ \wh R^\pr_{a b}, \ \wh D^{\pr b} \wh\psi^\pr_{b a+}),
\eea
which would be essential for discussing field equations in 2D supergravity.

We note that although the multiplets \eqref{Scalara-} to \eqref{Ma-} were obtained here by the action of supercovariant derivatives, the results obtained are independent of this fact. This can be checked by directly verifying that the supersymmetry algebra closes in the usual way, for the transformation rules given. 

Finally, we note that given a multiplet (\ref{Ma-}), we can form another multiplet 
\be
\label{star}
\star(\z_{a-},\ U_{ab},\ \l_{a+})
= ( \vare_{ab} \z^b_- ,\quad  U_{b a} - \h_{a b} U_c^{~c},\quad - \g_{a b} \l^b_+ -\g_{a b} \g^c \wh D_c^\pr \z^b_-  ). 
\ee
This `star transformation' appears in the boundary condition (\ref{SMBC}). Note that it squares to unity: $\star^2=1$.

\section{Irreducible submultiplets of reducible multiplets} \label{RedSplitting}

The reducible multiplet \eqref{Ma-} can be split into irreducible submultiplets. To do this, we begin by defining the projection tensors,
\bea
P_{\pm a b} \equiv \half(\eta_{a b}\pm\eps_{a b}),
\eea
which enjoy the following properties
\bea
&& P_{+a b}+P_{-a b}=\eta_{a b}, \quad P_{+ a b}=P_{-b a}, \nn\\
&& P_{\pm a}{}^b P_{\pm b}{}^c=P_{\pm a}{}^c, \quad
P_{\pm a}{}^b P_{\mp b}{}^c=0, \quad
P_{\pm}{}^{a b}P_{\pm a b}=0.
\eea
Using these projection operators, we find that $(\z_{a-}, U_{ab}, \l_{a+})$ contains the following irreducible submultiplets
\ba
\label{irreps}
\P_2(\g^a \z_{a-}) &=  \Big(\ga^a\z_{a-}, \quad 2P_{-}{}^{a b}U_{a b}\Big), \nn\\
\P_2(\ga_a\ga^c\ga_b\z_{c-}) &= \Big(\ga_a\ga^c\ga_b\z_{c-}, \quad 4P_{+a}{}^c P_{+b}{}^d U_{c d}\Big) ,\nn\\
\F_2(2P_{+}{}^{a b}U_{a b} ) &= \Big(2P_{+}{}^{a b}U_{a b}, \quad 2\ga^a\la_{a+}+\ga^a\ga^b\inD_b\z_{a-}\Big) ,\nn\\
\F_2( 4P_{-a}{}^c P_{-b}{}^d U_{c d} ) &= \Big(4P_{-a}{}^c P_{-b}{}^d U_{c d}, \quad 2\ga_a\ga^c\ga_b\la_{c+}+\ga_a\ga^c\ga_b\ga^d \inD_c \z_{d-}\Big), 
\ea
transforming as $(\chi_{+},F)$, $(\chi_{a b+},F_{a b})$, $(A,\chi_{-})$ and $(A_{a b},\c_{a b-})$ multiplets, respectively. 

These multiplets can alternatively be expressed in terms of light-cone coordinates,~\footnote{
The relationship between the projection tensor $P_{\pm ab}$ and the light-cone coordinates can be highlighted by noting that on a 2D vector $v_m=(v_0,v_1)^T$, the projection tensor acts as
\ba
2P_{+a}{}^{b} v_b &= \left( 
\begin{array}{c}
v_\pl \\
v_\pl
\end{array}
\right), &
2P_{-a}{}^{b} v_b &=\left( 
\begin{array}{c}
v_\mi \\
- v_\mi
\end{array}
\right) , \nn
\ea
where $v_{\plmi}=v_0\pm v_1$.
From this it is clear that the action of the projection tensor produces a vector parametrized by one light-cone coordinate element.
} 
where we define $\pa_\plmi = \pa_{0} \pm \pa_{1} $ and $\z_{\plmi - } = \z_{0 -} \pm \z_{1 -}$. We find, respectively,
\ba
&\Big( \g_\pl \z_{\mi -} , 2 U_{\pl \mi} \Big),  \nn\\
& \Big( \g_{\pl} \z_{\pl -}, 2 U_{\pl \pl} \Big), \nn\\
&\Big( U_{\mi \pl} , \g_{\mi} \l_{\pl +} + \inD_{\mi} \z_{\pl -} \Big), \nn\\
& \Big( U_{\mi \mi} , \g_{\mi} \l_{\mi +}  + \inD_{\mi} \z_{\mi - } \Big).
\ea
It is clear that each of these multiplets contains one component $U_{ab}$. The first two transform as $(\chi_{+},F)$, and the last two as $(A,\chi_{-})$ multiplets.

A similar splitting exists also for the multiplet \eqref{Mred}.


\end{document}